\documentclass[prb,showpacs,preprint,superscriptaddress]{revtex4}
\usepackage{amssymb,graphicx,amsbsy,amsmath}

\begin{document}

\title{Kosterlitz-Thouless transition of magnetic dipoles on the
two-dimensional plane}
\author{Seung Ki Baek}
\affiliation{Integrated Science Laboratory, Department of Physics, Ume{\aa}
University, 901 87 Ume{\aa}, Sweden}
\author{Petter Minnhagen}
\affiliation{Integrated Science Laboratory, Department of Physics, Ume{\aa}
University, 901 87 Ume{\aa}, Sweden}
\author{Beom Jun Kim}
\email[Corresponding author, E-mail: ]{beomjun@skku.edu}
\affiliation{BK21 Physics Research Division and Department of Physics,
Sungkyunkwan University, Suwon 440-746, Republic of Korea}
\affiliation{Asia Pacific Center for Theoretical Physics, Pohang 790-784,
Republic of Korea}

\begin{abstract}
The universality class of a phase transition is often determined by factors
like dimensionality and inherent symmetry. We study the magnetic
dipole system in which the ground-state symmetry and the underlying lattice
structure are coupled to each other in an intricate way. A two-dimensional
(2D)
square-lattice system of magnetic dipoles undergoes an order-disorder phase
transition belonging to the 2D Ising universality class.
According to Prakash and Henley [Phys. Rev. B {\bf 42}, 6572 (1990)], this can be
related to the fourfold-symmetric ground states which suggests a similarity
to the four-state clock model. Provided that this type of
symmetry connection holds true, the magnetic dipoles on a honeycomb lattice,
which possess sixfold-symmetric ground states, should exhibit a
Kosterlitz-Thouless transition in accordance with the six-state clock model.
This is verified through numerical simulations in the present investigation.
However, it is pointed out that this symmetry argument does not always
apply, which suggests that factors other than symmetry can be decisive for
the universality class of the magnetic dipole system.
\end{abstract}

\maketitle

Understanding the physics in thin films is of practical 
importance since thin-film construction is used in manufacturing
a variety of electronic and optical devices. One may note that magnetic thin
films, in particular, play a key role in massive data storage applications
and that magnetic interactions in such films will become more prominent as
magnetic moments are more densely integrated on such devices.
A typical behavior of the magnetic property in thin films is switching
between perpendicular and in-plane magnetization as the temperature $T$ varies,
which was first reported for Fe films.~\cite{pappas} On the other hand, for
rare-earth compounds such as ErBa$_2$Cu$_3$O$_{6+x}$, the ordering of magnetic
spins at $T < 1.0$K is essentially two-dimensional (2D).~\cite{lynn} For
these materials, the exchange interaction is known to be relatively
weak,~\cite{mac} and the rare-earth ionic moments have been described as
Ising or $XY$ spins depending on anisotropy.~\cite{simizu}
These compounds have also attracted
attention as a suitable candidate to reveal the relation between
superconductivity and magnetism. A model of a magnetic thin film is
therefore a
2D lattice of magnets governed by the dipole interaction and
confined to rotate on the plane of the lattice.
Although it is straightforward to write down the corresponding Hamiltonian,
understanding its physics is more cumbersome, for two reasons:
the long-range character of the dipole interaction and its anisotropy.
For a lattice of $N$ planar magnets with the dipole interaction, the
Hamiltonian is given as follows:
\begin{equation}
H = J \sum_{i \neq j} \left[ (\boldsymbol{s}_i \cdot \boldsymbol{s}_j)
r_{ij}^2 - 3 (\boldsymbol{s}_i \cdot \boldsymbol{r}_{ij}) (\boldsymbol{s}_j
\cdot \boldsymbol{r}_{ij}) \right] / r_{ij}^5,
\label{eq:h}
\end{equation}
where $J(>0)$ is a coupling constant and the summation runs over all the
distinct spin pairs $\boldsymbol{s}_i$ and $\boldsymbol{s}_j$, residing at
$\boldsymbol{r}_i$ and $\boldsymbol{r}_j$, respectively.
The distance between members of a spin pair is denoted as $r_{ij} = \left|
\boldsymbol{r}_i - \boldsymbol{r}_j \right|$.
Since we employ the periodic-boundary condition, the displacement
$\boldsymbol{r}_{ij}$ between $\boldsymbol{r}_i$ and $\boldsymbol{r}_j$
is chosen as the one with the minimal distance among
every possible pair of their periodic images. If more than two
periodic images of a spin have the same minimal distance from another spin,
we neglect the interaction between members of this spin pair to remove the ambiguity.
As is clearly seen in
Eq.~(\ref{eq:h}), the interaction energy between $\boldsymbol{s}_i$ and
$\boldsymbol{s}_j$ decays as $r_{ij}^{-3}$ and it does not depend solely on
their angular difference but also on their relative position,
$\boldsymbol{r}_{ij}$.
In terms of numerical analysis, the long-range character imposes an $O(N^2)$
complexity within the simple Metropolis algorithm and the anisotropy puts an
obstacle to developing an effective cluster algorithm. These problems have left the
properties of the phase transition in this system largely inconclusive.
On the one hand, this dipole lattice is related to the ice model,~\cite{bark}
which undergoes a phase transition keeping its structural arrangement
disordered.~\cite{bram} The lack of a long-range order has been reported in
the neutron-scattering experiments  for some rare-earth compounds~\cite{clinton}
and also in artificial spin ice.~\cite{wang} On the other hand,
such disorder-preserving behavior apparently disagrees with theoretical
and numerical predictions that the long-range order will be established in the
2D dipole lattice at low $T$.~\cite{debell1,debell2,car} Even if we
accept the existence of an order-disorder transition at a critical
temperature $T_c$ in the dipole system,
numerical studies produce conflicting results: For the case of a
square lattice, there has been reported a value of the critical exponent
$\beta=0.19(4)$ for the staggered magnetization as well as $\gamma =
1.37(7)$ for the staggered susceptibility with the correlation-length
exponent $\nu=1$,~\cite{car,ras} where the numbers in the parentheses
are numerical errors in the last digits. The 
staggered magnetization vector is defined as $\boldsymbol{m} = N^{-1} \sum_i
\boldsymbol{\sigma}_i$, where we specify the components of the vectors as
$\boldsymbol{r}_i = (x_i, y_i)$ and $\boldsymbol{s}_i = (\cos\theta_i,
\sin\theta_i)$, respectively, and define gauge-transformed spins as~\cite{debell1}
\begin{equation}
\boldsymbol{\sigma}_i \equiv \left[
(-1)^{y_i} \cos\theta_i, (-1)^{x_i} \sin\theta_i \right].
\label{eq:gauge}
\end{equation}
We take the magnitude $m = \left| \boldsymbol{m}
\right|$ as a scalar magnetic-order parameter of this system.
Using the same observable, a recent study
reported $\beta/\nu=0.13(2)$ and $\nu=1.05(5)$ with the Metropolis
algorithm.~\cite{fer} This result indicates the 2D Ising universality class within
errors, which is partially supported by another study on the Heisenberg
dipole system.~\cite{tomita} However, none of these
match a renormalization-group calculation~\cite{maier}
yielding an exponentially diverging correlation length $\xi$ with $\log \xi
\sim 1/\sqrt{T-T_c}$ and magnetization $\sim \xi^{-1/2}$.
To our knowledge, no theoretical explanation of the observed results has
been satisfactorily provided.

In this brief report, we begin with the critical behavior of magnetic
dipoles on the square lattice. The finite-size scaling result from $L \times
L$ square lattices supports an order-disorder transition with the 2D Ising
universality class, confirming the recent observation.~\cite{fer}
We then ask if this is related to the fact
that the system possesses fourfold-symmetric ground states.~\cite{debell1}
To test this further, we use the fact that the magnetic dipoles may have
sixfold-symmetric ground states if put on the honeycomb lattice.~\cite{pra}
This has been well established for the nearest-neighbor dipole
interaction,~\cite{pra} and we have numerically
checked its validity for the long-range-interaction case as well. We find
that the honeycomb-lattice lattice case exhibits a
Kosterlitz-Thouless (KT) transition as implied by the analogy to the
six-state clock model.~\cite{six,nonkt}

\begin{figure}
\begin{center}
\includegraphics[width=0.45\textwidth]{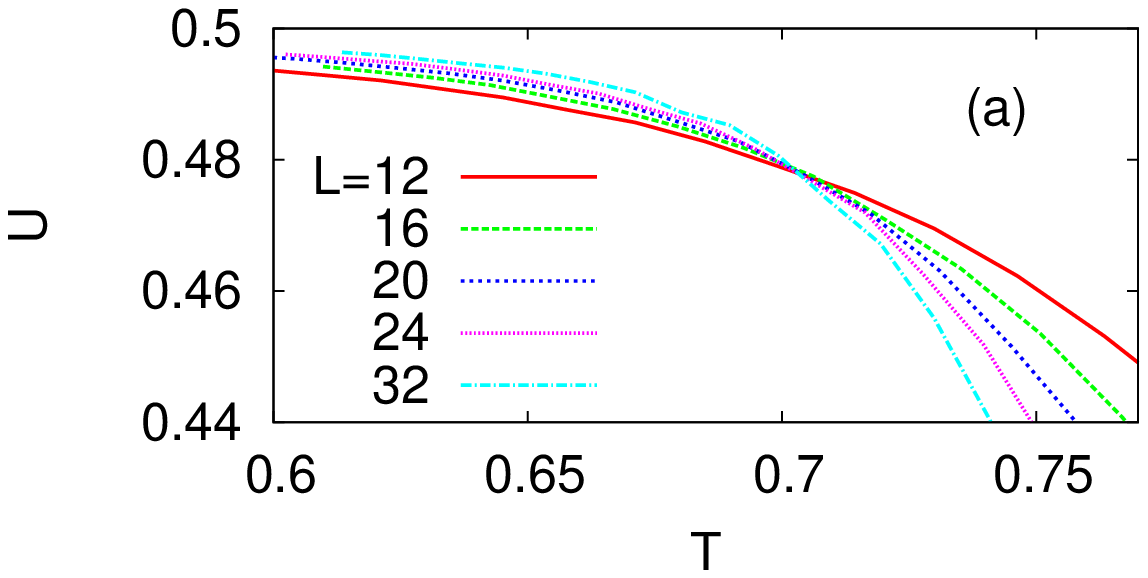}
\includegraphics[width=0.45\textwidth]{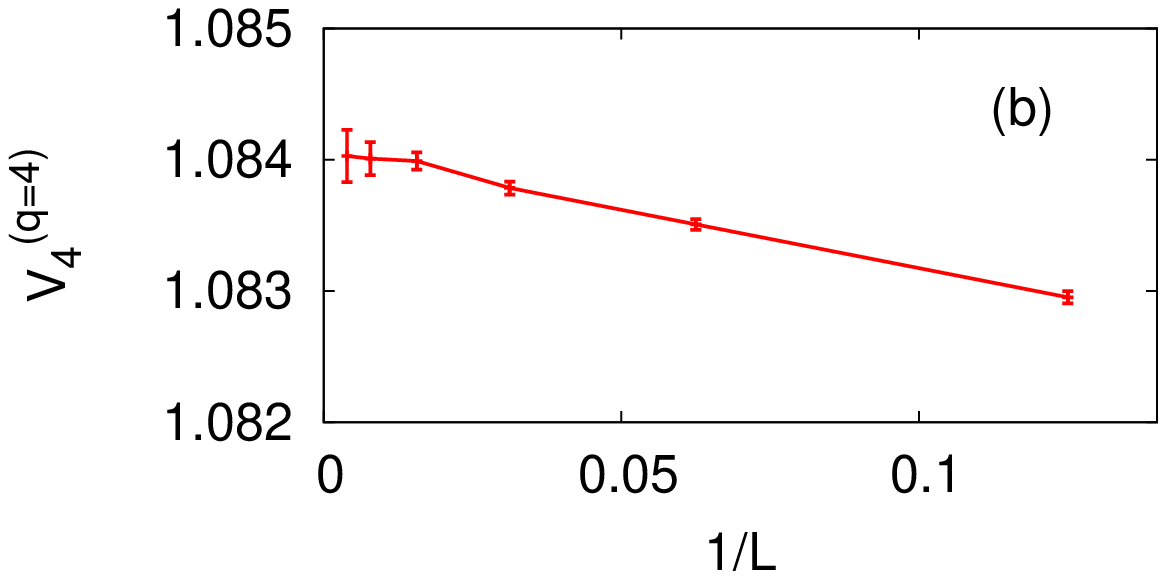}
\end{center}
\caption{(Color online) (a) Binder's cumulant of magnetic dipoles
on $L \times L$ square lattices.
(b) Cumulant ratio $V_4^{q=4} = \left<m^4\right> / \left<m^2\right>^2$
obtained for the 2D square-lattice four-state clock model at the critical
temperature $T_c^{q=4} = 1/\log(1+\sqrt{2})$.}
\label{fig:binder}
\end{figure}

Our simulation strategy is as follows: we use the parallel tempering (PT)
method that was devised to equilibrate glassy spin systems with very
long relaxation times.~\cite{exchange,mc} Consider simulating two samples of
a spin system in parallel at different inverse temperatures, $\beta_1$ and
$\beta_2$, respectively. If $\beta_1 < \beta_2$, the sample at $\beta_1$
will explore a larger region in the phase space than will the sample at
$\beta_2$. If the exploration happens to find a state with sufficiently low
energy, we exchange $\beta_1$ and $\beta_2$ of the two samples so that the
low-energy state can be pursued more deeply while the other sample begins a
new exploration. Specifically, if energies of the two samples are denoted
$E_1$ and $E_2$, respectively, the exchanging probability is given as
$P_{\rm ex} = \min \left\{ 1, \exp \left[ - (\beta_2 - \beta_1) (E_1-E_2)
\right] \right\}$ to satisfy the detailed balance. One may easily extend
this scheme to more than two samples, ranging over a broad temperature
region, and run the samples simultaneously on parallel computing devices. We
simulate each sample by the Metropolis algorithm, which means that the
overall complexity is still $O(N^2)$. Only its proportionality coefficient
will be reduced by application of the PT method, but it is nevertheless a
significant gain in practice especially when $N$ is not too large. For
$L=32$ in Fig.~\ref{fig:binder}(a), for example, it usually takes $10^4 \sim
10^5$ Monte Carlo steps for a simple Metropolis algorithm to equilibrate the
system around $T_c$ from a random configuration, while it is enough to make
a couple of exchange moves with running $O(10^3)$ steps in between. We
determine the difference in inverse temperature by observing overlaps of
energy histograms. In the same example ($L=32$), we have simultaneously
simulated $31$ inverse temperatures over $[1.03,1.63]$, each of which runs
on an Intel Xeon quad-core L5420 CPU (2.5GHz). The CPU time spent for this
size is about $6\times 10^2$ h per temperature, meaning that
the total CPU time to obtain the result for $L=32$ roughly amounts to
$2\times 10^4$ h.
 
Figure~\ref{fig:binder}(a) shows Binder's cumulant
$U \equiv 1 - \frac{1}{2} \left< m^4 \right>/\left< m^2 \right>^2$
from the staggered magnetization, where $\left< \cdots \right>$ means
thermal average. Note that this cumulant is scaled to approach zero at high
$T$ and $1/2$ at low $T$ since
$\left<m^4\right> = 2\left<m^2\right>^2$ when
the magnetization vector $\boldsymbol{m}$ has a 2D Gaussian distribution
centered at the origin.
From Fig.~\ref{fig:binder}(a),
the transition temperature is estimated as $T_c=0.70(2)$ in units of
$J/k_B$,
where $k_B$ is the Boltzmann constant. Since this is not an extremely precise
estimation, it is hard to get critical exponents to a good precision.
Instead, we may check consistency by assuming the 2D Ising universality
class.
\begin{figure}
\includegraphics[width=0.45\textwidth]{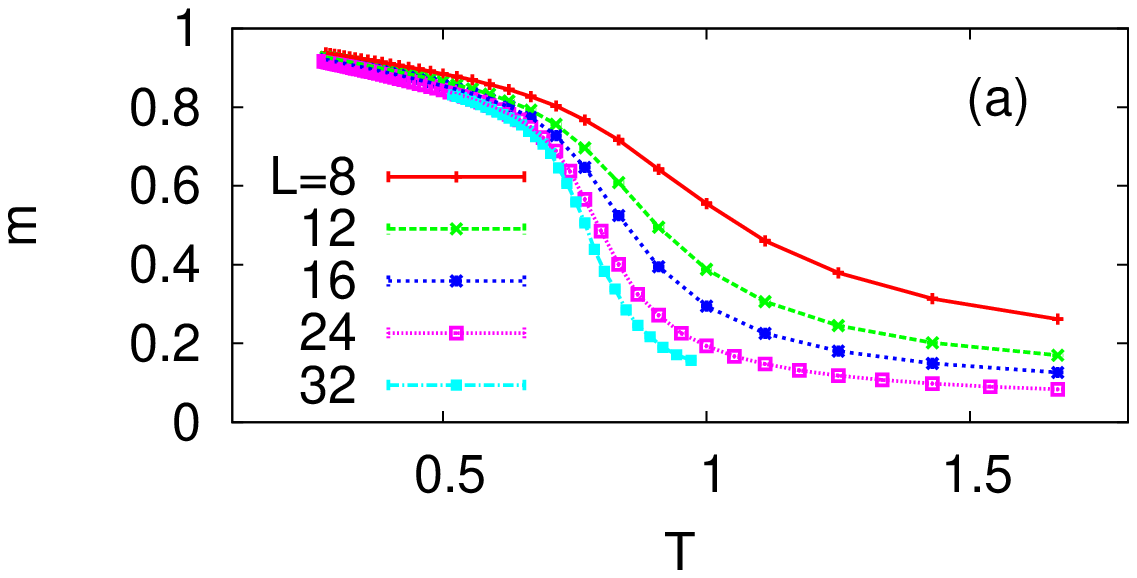}
\includegraphics[width=0.45\textwidth]{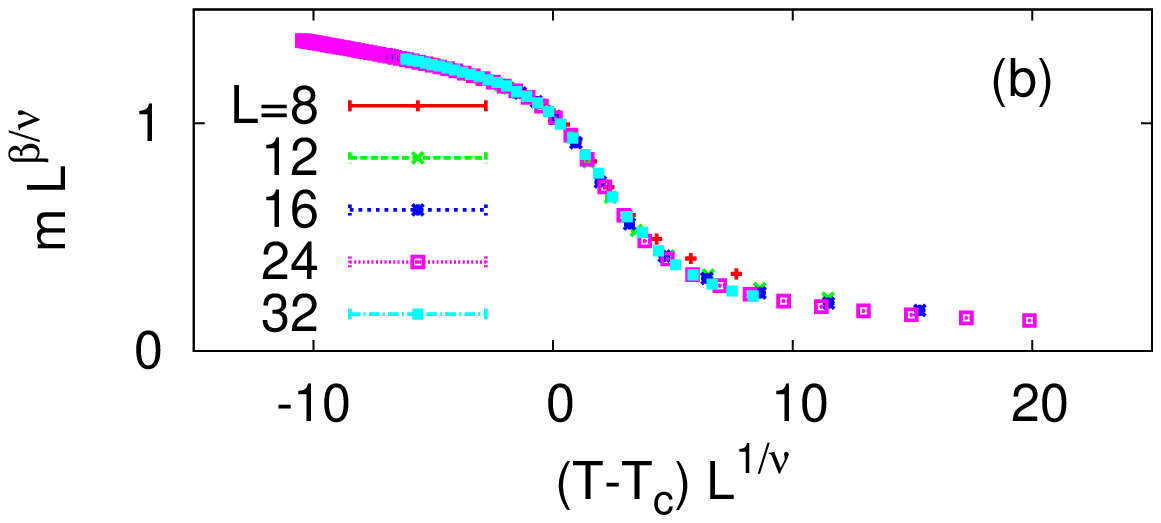}\\
\includegraphics[width=0.45\textwidth]{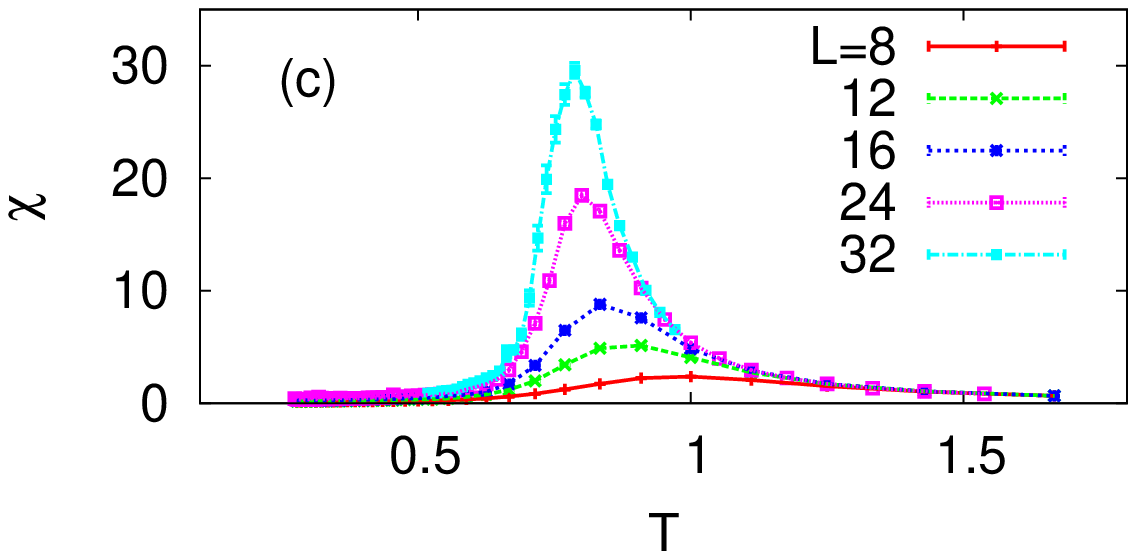}
\includegraphics[width=0.45\textwidth]{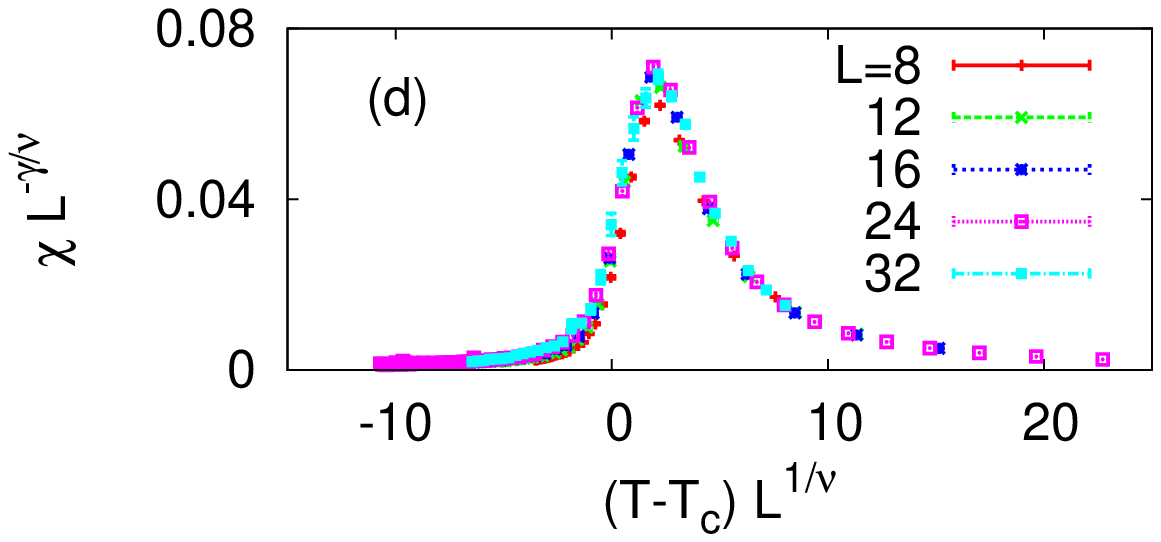}
\caption{(Color online) Results on square lattices. (a) Staggered
magnetization and its scaling collapse (b) according to the 2D Ising
universality class. (c) Staggered susceptibility and its scaling
collapse (d) in the same way. The 2D Ising critical exponents $\beta = 1/8$,
$\nu = 1$, and $\gamma = 7/4$ are used; and the critical temperature is
estimated to be $T_c = 0.71(1)$ and $0.72(1)$ for (b) and (d),
respectively.}
\label{fig:obs}
\end{figure}
Let us plot the staggered magnetization and try a scaling collapse by
the 2D Ising universality class, i.e., $\beta = 1/8$ and $\nu = 1$
[Figs.~\ref{fig:obs}(a) and \ref{fig:obs}(b)]. The best
collapse is observed at $T_c = 0.71(1)$, which is in good 
agreement with the value of $T_c$ estimated above via Binder's cumulant.
Another important observable is the staggered susceptibility
$\chi = N T^{-1} \left[ \left<m^2\right> - \left< m \right>^2 \right]$
plotted in Fig.~\ref{fig:obs}(c). Again the 2D Ising universality class with
the corresponding critical exponent $\gamma=7/4$ is clearly consistent with the data provided that $T_c = 0.72(1)$ [Fig.~\ref{fig:obs}(d)].
One also notes that the susceptibility data imply a single transition
point where the critical fluctuation diverges in the thermodynamic limit,
and that the susceptibility remains finite below this divergence. This
rules out the possibility of a KT transition. The
consistent descriptions strongly suggest that the critical behavior
indeed belongs to the 2D Ising universality class.

It is obviously nontrivial that a continuous-spin system, complicated by
the long-range character and anisotropy, nevertheless displays an Ising
transition.
The most plausible explanation is that the ground states of the dipole system
on the square lattice possess a fourfold symmetry,~\cite{debell1} since
the same is true for the four-state clock model, which exhibits the 2D Ising
criticality.~\cite{suzuki}
The value of the cumulant indeed tells us more than this simple symmetry
argument: Figure~\ref{fig:binder}(a) shows that $V_4 \equiv
\left<m^4\right>/\left<m^2\right>^2 = 2(1-U)
\approx 1.04(1)$. For the 2D Ising model, on
the other hand, the value of this quantity is given as $V_4^{\rm Ising} =
1+\epsilon$, where $\epsilon = 0.167923(5)$.~\cite{salas} Since the four-state
clock model is equivalent to two independent Ising systems $A$ and $B$ with
the temperature rescaled,~\cite{suzuki} we can denote it as $A \otimes B$
and consider its magnetic-order parameter 
$m = \sqrt{(m_A^2 + m_B^2)/2}$ where $m_A$ and $m_B$ correspond to
magnetizations of the independent Ising systems. This leads to
the cumulant value of the four-state clock model as follows:
\begin{eqnarray*}
V_4^{(q=4)} &=& \frac{\left< (m_A^2 + m_B^2)^2 \right>}{\left< m_A^2 + m_B^2
\right>^2} = \frac{\left<m_A^4\right> + 2\left< m_A^2 \right> \left< m_B^2
\right> + \left< m_B^4 \right>}{\left< m_A^2 \right>^2 + 2\left< m_A^2
\right> \left< m_B^2 \right> + \left< m_B^2 \right>^2}\\
&=& \frac{2 + 2V_4^{\rm (Ising)}}{4} = 1 + \frac{\epsilon}{2},
\end{eqnarray*}
which agrees well with our Monte Carlo calculation
$V_4^{q=4} = 1.0840(2)$ at $T_c^{q=4} = 1/\log(1+\sqrt{2})$
[Fig.~\ref{fig:binder}(b)].
By the same analogy, the cumulant value for the
magnetic dipoles suggests that $V_4 = 1+\epsilon/4 \approx 1.042$, which
could be observed in a combination of four independent Ising systems denoted
as $A,B,C$, and $D$, respectively, or two independent four-state clock
models $A \otimes B$ and $C \otimes D$, where the total magnetization is
equivalent to $m = \sqrt{(m_A^2 + m_B^2 + m_C^2 + m_D^2)/4}$.

In order to test the symmetry argument further, we investigate the honeycomb
lattice, since the nearest-neighbor dipole interaction in this case leads
to sixfold ground states.~\cite{pra}
In addition, since it is known that the $q$-state clock model with the
cosine interaction undergoes a KT transition when $q \ge
6$,~\cite{six,nonkt} the symmetry argument suggests that the transition for
a honeycomb lattice should be of a KT type.

\begin{figure}
\begin{center}
\includegraphics[width=0.48\textwidth]{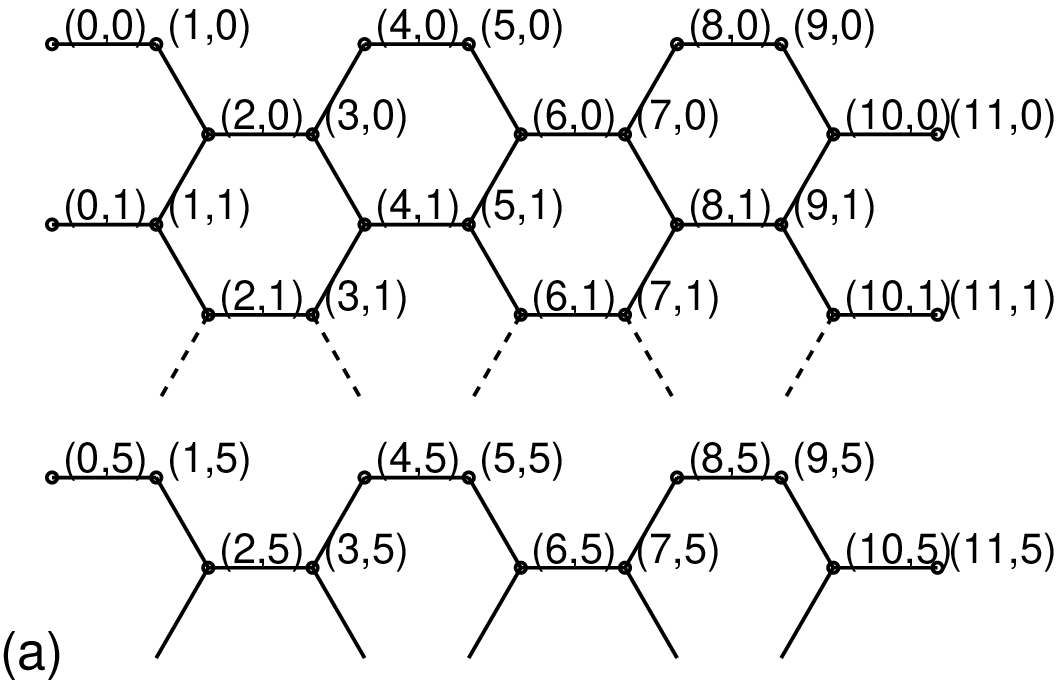}
\includegraphics[width=0.33\textwidth]{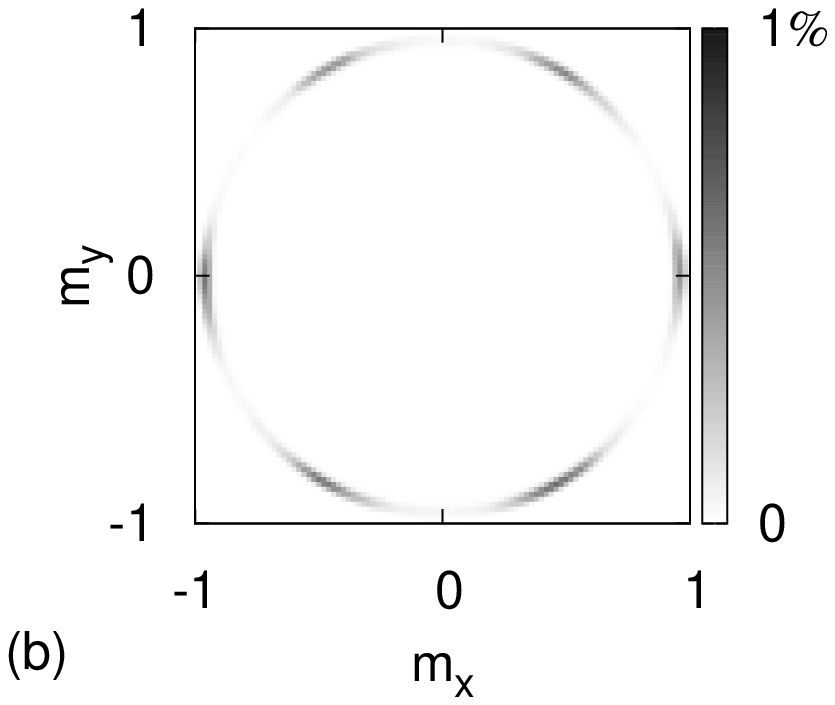}
\end{center}
\caption{(a) Honeycomb lattice of size $L=6$ with the periodic-boundary
condition. Coordinates are presented as $(u,w)$ for the gauge
transformation in Table~\ref{table:gauge}. (b) Distribution of staggered
magnetization $\boldsymbol{m} = (m_x, m_y)$, showing the sixfold symmetry,
taken at $T=0.1$ with $L=12$.
}
\label{fig:honey}
\end{figure}

\begin{figure}
\includegraphics[width=0.45\textwidth]{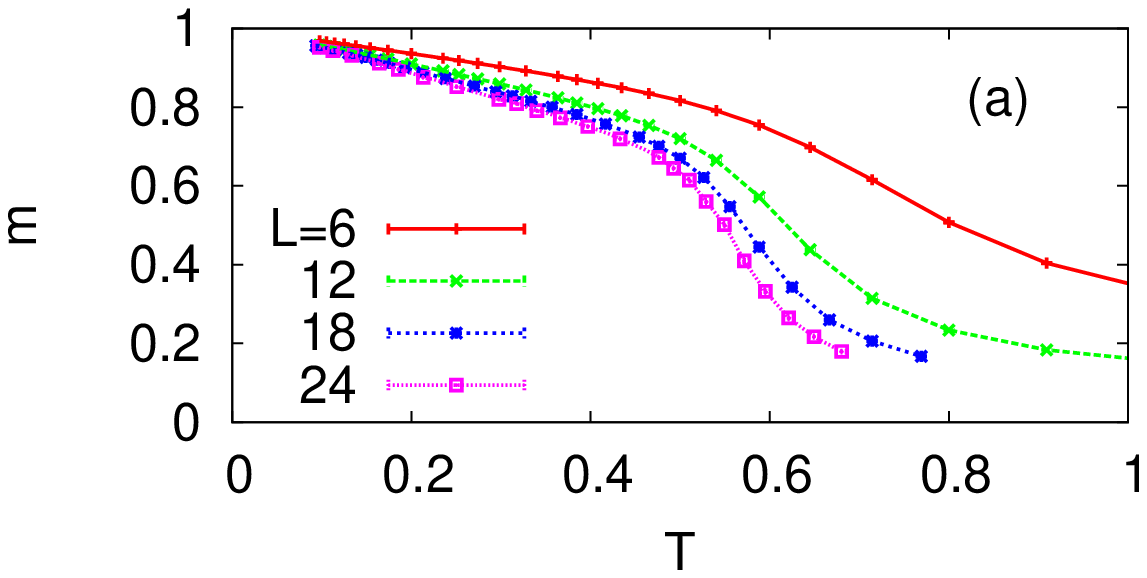}
\includegraphics[width=0.45\textwidth]{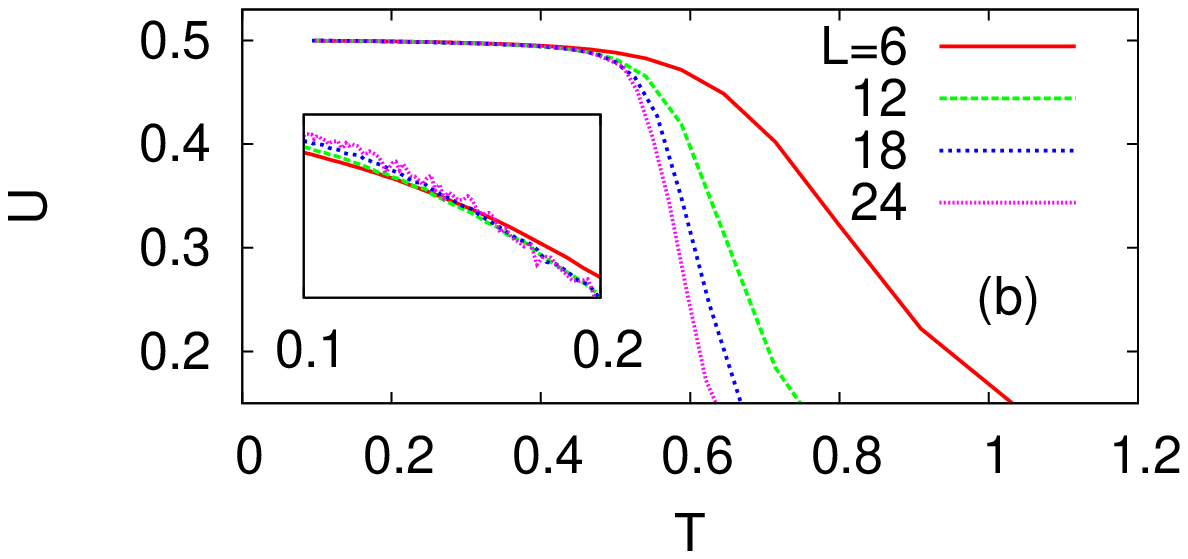}\\
\includegraphics[width=0.45\textwidth]{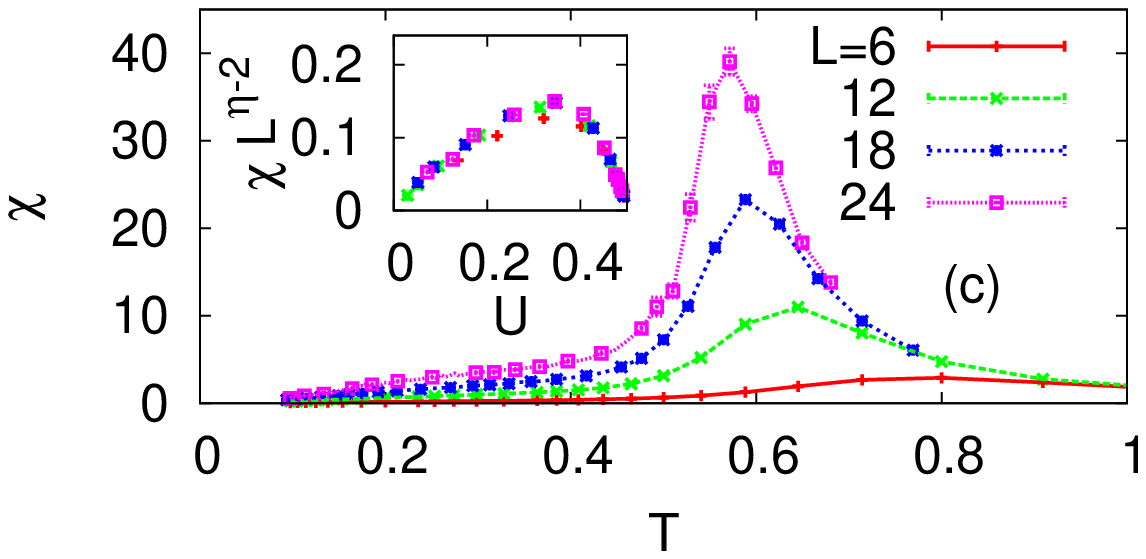}
\includegraphics[width=0.45\textwidth]{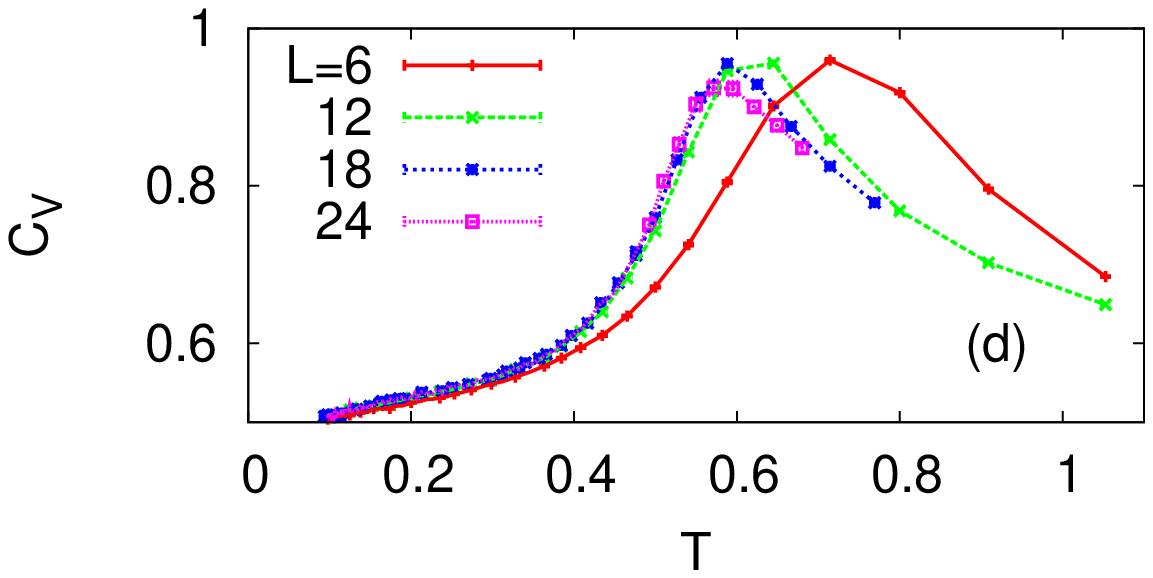}
\caption{(Color online) Results on honeycomb lattices. (a) Magnitude of the
staggered magnetization and (b) Binder's cumulant, with a closer view of the
low-$T$ part (inset). (c) Staggered susceptibility $\chi$ and the scaling
collapse of $\chi \sim L^{2-\eta}f(U)$ with $\eta = 1/4$. (d) Specific
heat does not show any clear double-peak structure.}
\label{fig:hh}
\end{figure}

The honeycomb lattice that we use in this work is shown in
Fig.~\ref{fig:honey}(a) where the system size is $N=2L^2$ for a given
length scale $L$. We choose $L$ as multiples of $6$ by taking the ground-state
configurations~\cite{pra} into consideration. 
An appropriate gauge transformation of the spin angle
$\theta$ at site $(u,w)$ in the same spirit of Eq.~(\ref{eq:gauge}) is
tabulated in Table~\ref{table:gauge}. The staggered magnetization is defined
as $\boldsymbol{m} = N^{-1} \sum (\cos \theta', \sin \theta')$
with magnitude $m \equiv |\boldsymbol{m}|$.
As in the square-lattice case, each chosen spin interacts with a set of other
spins that have well-defined minimal distances to the chosen spin.  We
furthermore require this interacting set to have the inherent symmetry of the
honeycomb lattice; i.e., the interacting set should be left invariant under
the rotation by $\pm 2\pi/3$ around the chosen spin so that
the six ground states are equally probable in the system
[Fig.~\ref{fig:honey}(b)].  The
numerical results are again obtained by the
PT method and plotted in Figs.~\ref{fig:hh}(a)-\ref{fig:hh}(d),
where Binder's cumulant
$U$ and the staggered susceptibility $\chi$ are given in the same way as above.
In order to examine the low-$T$ phase, we have the inverse temperature
range over $O(1)$ to $O(10)$ while keeping the overlaps in the
energy histograms. For $L=24$, for example, we ran $127$ inverse
temperatures over $[1.5,10.3]$ in parallel, spending about $10^2$ CPU hours
per each.
Around $T \approx 0.5$, one finds a size merging of $U$, together with a
divergence in $\chi$, which are characteristic signatures of the KT
transition. These observations imply the existence of a quasicritical phase
below $T \approx 0.5$ where the correlation length diverges. The KT picture
also predicts a scaling collapse of $\chi$ vs $U$, such that $\chi \sim
L^{2-\eta} f(U)$ where $f$ is a certain scaling function and $\eta =
1/4$.~\cite{gqclock} This method provides a piece of information about the
universality class even without precise knowledge of the transition
temperature. In spite of the small system sizes, the KT scaling exponent
$\eta=1/4$ nevertheless gives a scaling collapse consistent with a KT
transition as shown in the inset of Fig.~\ref{fig:hh}(c). We also note an
additional tiny yet systematic size dependence of $U$ below $T = 0.2$ shown
in the inset of Fig.~\ref{fig:hh}(b), which possibly indicates that the
staggered magnetization freezes into the sixfold symmetry [compare
Fig.~\ref{fig:honey}(b)]. Unlike in the six-state clock model, however,
this freezing is not accompanied by any peak in specific heat
[Fig.~\ref{fig:hh}(d)], which implies that the quasicritical phase is
not isotropic either but should reflect the sixfold symmetry at
least in part. It is currently under investigation whether the $U(1)$
symmetry in the disordered phase gets broken exactly at the same
temperature where the KT transition occurs.

\begin{table}
\caption{Gauge transformation for the honeycomb lattice, where $(u,w)$
represents the coordinates as shown in Fig.~\ref{fig:honey}(a).}
\begin{tabular*}{\hsize}{@{\extracolsep{\fill}}ccc|ccc|ccc}\hline\hline
$u \mod 4$ & $w \mod 3$ & $\theta'$ &
$u \mod 4$ & $w \mod 3$ & $\theta'$ &
$u \mod 4$ & $w \mod 3$ & $\theta'$ \\\hline
0 & 0 & $\theta$ &
0 & 1 & $\theta + 2\pi/3$ &
0 & 2 & $\theta - 2\pi/3$\\
1 & 0 & $-\theta$ &
1 & 1 & $-\theta + 2\pi/3$ &
1 & 2 & $-\theta - 2\pi/3$ \\
2 & 0 & $\theta - 2\pi/3$ &
2 & 1 & $\theta$ &
2 & 2 & $\theta + 2\pi/3$\\
3 & 0 & $-\theta - 2\pi/3$ &
3 & 1 & $-\theta$ &
3 & 2 & $-\theta + 2\pi/3$ \\\hline\hline
\end{tabular*}
\label{table:gauge}
\end{table}

In summary, we have confirmed that the $XY$-type magnetic dipoles on the
square lattice exhibit the 2D Ising criticality and that this can
be related to a symmetry similarity in the four-state clock model.
This symmetry connection is further supported
by the study of the honeycomb lattice, where the ground states have sixfold
symmetry;
and the system behaves similarly to the six-state clock model
exhibiting a KT transition.
However, the symmetry argument is not always the decisive factor: the
transition is KT-like for the square lattice with nearest-neighbor
interaction in spite of the fact that the symmetry remains the
same.~\cite{romano}
One may also note that the long-range order at low $T$ is also
absent in experiments with squarelike structures.~\cite{clinton} However,
the factors that supersede the symmetry argument in deciding the critical
universality class remain to be elucidated.
 
\acknowledgments
S.K.B. and P.M. acknowledge support from the Swedish Research Council
with Grant No. 621-2008-4449; 
B.J.K. was supported by the Priority Research Centers Program through the National 
Research Foundation of Korea (NRF) funded by the Ministry of Education,
Science and Technology (2010-0029700).
This research was conducted using the resources of High Performance
Computing Center North (HPC2N).


\end{document}